\documentclass[11pt, final]{myarticle} 
\usepackage{setspace}
\usepackage{amssymb,cite,latexsym,amsmath,subfigure}
\usepackage{graphicx}
\usepackage{psfrag}
\usepackage{pstricks}
\makeatletter
\renewcommand\@seccntformat[1]{\csname the#1\endcsname.\quad}

\date{}
\begin{document} 

\title{{\Large 
Stochastic solution of nonlinear and nonhomogeneous
evolution problems by a differential Kolmogorov equation}}
\author{R. G. Keanini\footnote{
Address: Mechanical Engineering \& Engineering Science, UNC-Charlotte,
9201 University City Blvd, Charlotte, NC 28223-0001.
Email: rkeanini@uncc.edu; Phone: 704-687-8336; Fax: 704-687-8345}\\
Department of Mechanical Engineering and Engineering Science\\
The University of North Carolina at Charlotte \\
Charlotte, North Carolina 28223-0001}
\maketitle
\section*{Abstract}
A large class
of physically important nonlinear and nonhomogeneous 
evolution problems,
characterized by advection-like and diffusion-like processes,
can be usefully studied by a time-differential form of
Kolmogorov's solution of the backward-time Fokker-Planck equation.
The differential solution embodies an integral representation theorem
by which any physical or mathematical entity satisfying a 
generalized nonhomogeneous advection-diffusion equation
can be calculated incrementally in time.
The utility of the approach for tackling nonlinear
problems is illustrated via solution of the noise-free
Burgers and related Kardar-Parisi-Zhang (KPZ) equations where it
is shown that the differential Kolmogorov solution 
encompasses, and allows derivation of, the 
classical Cole-Hopf and KPZ transformations and solutions. 
A second example, illustrating application of 
this approach to nonhomogeneous
evolution problems, derives the Feynman-Kac formula 
appropriate to a Schr\"odinger-like equation.

\vspace{.4cm}

\noindent \textbf{Key words.} nonlinear partial differential equations,
stochastic methods, Cole-Hopf transformation, Feynman-Kac formula

\vspace{.4cm}

\noindent \textbf{AMS subject classifications.} 82C31, 35K55, 60H30

\vspace{.4cm}

\noindent \textbf{Abbreviated title.} Stochastic solution of evolution equations

\vspace{.8cm}

\noindent \textbf{1. Introduction} \\
A wide array of physically important time-evolutionary problems, in areas
as diverse as hydrodynamic turbulence \cite{bouchand, she, kraichnan}, quantum mechanics
\cite{shankar,kaniadakis}, and the evolution of cosmic large-scale structure 
\cite{vergassola}, 
can be modeled by nonlinear equations of the generic form
\begin{equation}\label{general}
\eta_{\theta} + \mathbf{b} \cdot \nabla \eta + \nu \nabla^2 \eta = f(\eta,\eta_i,\eta_{ij \dots n})
\end{equation}
where $ \eta $ represents a scalar field 
whose evolution within a finite or non-finite
space-time domain is determined, at least in part, by
processes that in some fashion, are analogous
to advection and/or molecular diffusion within a hydrodynamic flow.
Here, $ \mathbf{b} $ and
$ \nu $ correspond to, or are analogs 
of, the local dimensional or nondimensional 
drift field and diffusion coefficient, respectively,
while the source term, $ f , $ can include linear and nonlinear terms in both
$ \eta $ and derivatives of $ \eta $ up to order $ n . $ 
In addition, since we will conceptualize the incremental, stochastic construction of 
$ \eta $ as a procedure in which
random walkers are launched from a space-time solution point
$ ( \mathbf{x}, \theta ) $ \textit{toward} known/previously calculated $ \eta , $
we will follow Kolmogorov and 
state the evolution of $ \eta $ in terms of
backward time, $ \theta . $  

It is well known, particularly in the applied mathematics community \cite{friedman, schuss}
that \textit{linear} models, which are encompassed by (\ref{general}),
can be solved over finite (non-incremental) time intervals via
application of Kolmogorov's solution of the  
backward-time Fokker Planck (FP) equation.
In cases where (\ref{general}) has either 
a nonlinear or linear source term, $ f , $ in $ \eta , $
or a drift $ \mathbf{b} $ and/or diffusivity 
dependent on $ \eta $
(where the latter can arise in more general forms
of the third term on the left of (\ref{general})),
a Kolmogorov solution can still be \textit{expressed}
in representative or schematic form \cite{ma, pardoux}; however, in
these instances, one or more 
stochastic path integrals in the \textit{a priori}
unknown scalar field, $ \eta ( \mathbf{x}, \theta ) , $ appear.
While one can, in principal, formulate a system of 
equations in the unknown $ \eta , $ 
and attempt a 
numerical solution over the entire space-time domain, $ Q , $
this is computationally expensive and apparently
remains untried. Here, we focus on
simpler, low cost, and
often analytically tractable
local solutions.

The purpose of this paper is to point up the utility
of applying a \textit{time-differential} version of
Kolmogorov's solution to nonlinear and nonhomogeneous evolution problems of  
the form in (\ref{general}).
In order to demonstrate the potential of this approach for tackling nonlinear
problems, we show how the differential Kolmogorov solution can be
used to derive the Cole-Hopf and related Kardar-Parisi-Zhang (KPZ) solutions of the noise-free
Burger's \cite{burger, whitham} and KPZ
\cite{kardar} equations.  We then apply this approach 
to nonhomogeneous problems, deriving the Feynman-Kac formula as a 
solution to a Schr\"odinger-like equation.
We find that, at least in these test cases,
the differential Kolmogorov solution offers an alternative, relatively straightforward,
physically transparent basis for solution.
Moreover, since this approach can be adapted to accommodate
the effects of boundary interactions/boundary conditions on the evolution
of $ \eta $ \cite{keanini} (see also the caption to Fig. 2),
the potential for further applications seems clear.

Our jumping-off point
is thus Kolmogorov's solution of the backward-in-time Fokker-Planck equation,
connecting
the stochastic kinematics of random walkers 
within a finite or non-finite
space-time region, $ Q = D \times (\theta , T ] , $ to the evolution of 
the scalar field, $ \eta ( \mathbf{x}, \theta ) , $
within $ Q . $ Here, $ D $ is the spatial domain, $ \theta = T - t , $ is again
backward time, 
$ t $ forward time, and
$ T $ the forward time at which a solution for $ \eta  (\mathbf{x} , \theta ) $
is sought. The solution, 
written here for
finite 
$ Q , $ and represented schematically in Fig. 1, reads
\cite{friedman} :
\begin{equation}\label{sol1}
\eta (\mathbf{x}, \tilde{\theta}) = 
E_{ \mathbf{x}, \tilde{\theta}} \big[ g \big( \boldsymbol{\chi}(\tau) , \tau \big) \big]
+ E_{ \mathbf{x}, \tilde{\theta}} \big[ \phi \big( \boldsymbol{\chi} (\tau) \big) \big]
- E_{ \mathbf{x}, \tilde{\theta}} \big[ \int_{\tilde{\theta}}^{\tau}
f \big( \boldsymbol{\chi} (s), s \big) ds \big]
\end{equation}
where the first two terms on the right 
capture the effects of the random walker (RW) swarm, launched from the solution
point $ (\mathbf{x}, \tilde{\theta} ) , $
sampling (typically known) Dirichlet conditions,
$ \eta = g(\mathbf{x}, \theta' ) , $ on the boundary $ \delta Q $ of $ Q , $
and the final condition $ \eta = \phi (\mathbf{x} , \theta = T )  $
on the final time-slice, $ D \times \{ \theta = T \} . $
In cases where Neumann and/or mixed conditions are
imposed on $ \delta Q , $ an additional term similar to the first
appears \cite{keanini}. Throughout, $ E $ will represent the expectation
taken with respect to the solution point.
In addition, in (\ref{sol1}), $ \tau $ represents
the random time at which any given RW impinges either on 
$ \delta Q $ (first term on right side of (\ref{sol1})),
or on $ D \times \{\theta = T \} $ (second term).

\begin{figure}
\centering
\includegraphics[width=3.3in]{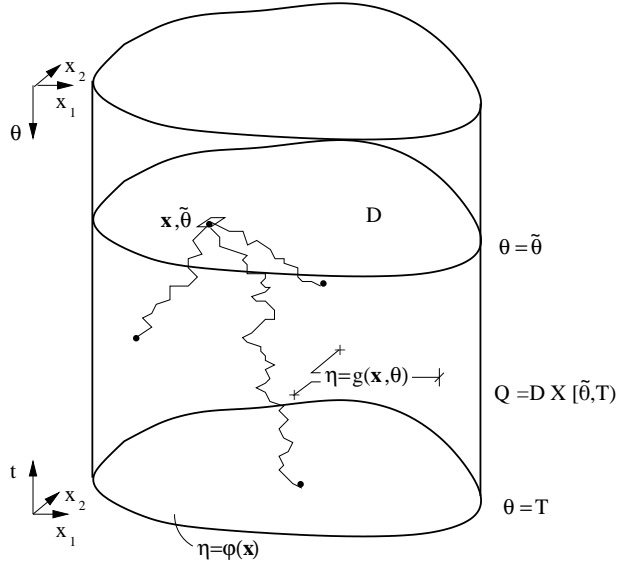}
\caption{Kolmogorov's local, non-time-incremental stochastic construction 
of the scalar $ \eta ( \mathbf{x}, \tilde{\theta} ) , $
as given by Eq. (\ref{sol1}),
appropriate to linear versions of Eq. (\ref{general}),
where the drift, $ \mathbf{b} , $ and diffusivity,
$ \nu $ (or their non-hydrodynamic analogs) are known.
In this illustration, 
$ \eta $ is subject to the Dirichlet condition $ \eta = g ( \mathbf{x} , \theta ) $
on the boundary $ \delta Q $ of the space-time cylinder $ Q , $
while on the final-time slice, $ D \times \{ \theta = T \} , $ 
$ \eta = \phi (\mathbf{x}) . $ }
\end{figure}

Couching the development
mainly in hydrodynamic terms, and depicting the time-incremental
solution in Fig. 2, 
we focus below on a differential 
form of (\ref{sol1}), expressed over 
short backward time intervals,
$ \Delta \theta = \theta_{j+1} - \theta_j , $ where $ \Delta \theta $
is chosen so that
the magnitudes 
of all three terms on the left side of (\ref{general}) remain comparable over the interval.
This is an important point both physically and computationally and
translates to the condition 
$ | ( \mathbf{x} - \mathbf{\bar{x}}(\theta_{j+1}) ) |^2 / l_D^2 
\approx
[b(x , \theta_j ) \Delta \theta]^2/ (\nu  \Delta \theta) = b l_b / \nu = O(1) , $
where $ b = | \mathbf{b}( \mathbf{x}, \theta ) | , $ 
$ \mathbf{b}(\mathbf{x} , \theta_j ) = - \mathbf{u}(\mathbf{x} , \theta_j ) $ is the local drift,
$ l_D = \sqrt{ \nu \Delta \theta } $ is the local diffusion length scale, 
$ l_b = b \Delta \theta $ is the local advection length scale,
and $ \mathbf{\bar{x}}(\theta_{j+1}) $
is the mean position of the random walk swarm at $ \theta_{j+1} . $ 
[Note, $ \mathbf{\bar{x}}(\theta_{j+1}) =
\mathbf{x} + E \int_{\theta_j}^{\theta_{j+1}}  
\mathbf{b} (\boldsymbol{\chi}(s') ds' = \mathbf{x} + 
\mathbf{b}(\mathbf{x}, \theta_j) \Delta \theta + O (\Delta \theta^2) ]. $

\begin{figure}
\centering
\includegraphics[width=3.3in]{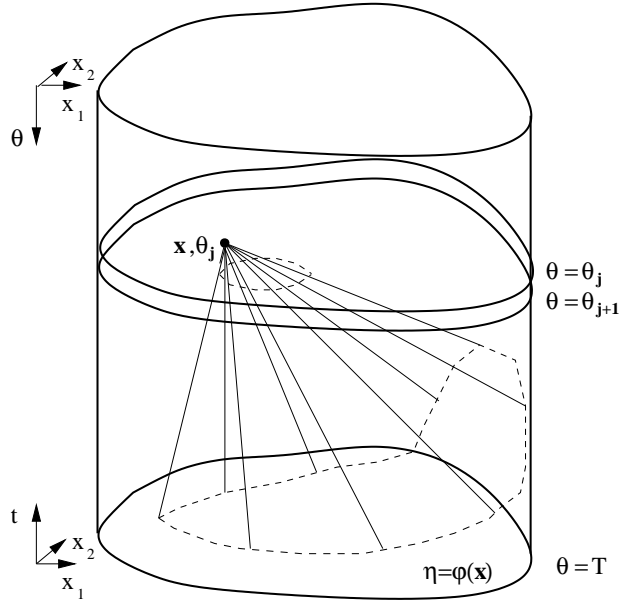}
\caption{Incremental stochastic construction 
of $ \eta ( \mathbf{x}, \theta_j ) , $
appropriate to nonlinear versions of Eq. (\ref{general}).
Here, an approximate 'cone-of-influence' is shown,
indicating how the local solution at $ (\mathbf{x} , \theta_j ) $ would
typically depend on previously determined $ \eta , $ 
computed incrementally over 
$ [ \theta_{j+1},T) . $  In this illustration,
a drift field (or drift field analog), $ \mathbf{b} , $ 
having a significant $ x_1 $
component, pushes RW's predominantly to the right.
Near Dirichlet boundaries, RW impingement with a boundary element
$ \delta Q_k = \delta D \times \Delta \theta_k $ (where $ \delta D $ is the boundary of
$ D $ and $ \Delta \theta_k = \theta_{k+1} - \theta_k ) , $
is readily accommodated using the first expectation in Eq. (\ref{sol1}), 
applied over $ \delta Q_k . $
A similar (or additional) term
would be required near Neumann (or mixed) boundaries \cite{keanini}.}
\end{figure}

Thus, in hydrodynamic terms, we arrive at the requirement that $ \Delta \theta $
must be chosen so that the local 
Reynolds number, $ Re = b l_b / \nu , $ remains order 1, or more meaningfully,
so that the local advection length scale, $ l_b = b \Delta \theta , $ is on the order of
the local diffusion length scale, $ l_D . $ 
More importantly, 
this condition implies that non-fixed
time steps may, under various circumstances, be required in both
analytical \cite{keanini} and Monte Carlo solutions,
a point recognized in the applied math literature \cite{milshtein},
but apparently not well known in the engineering and physics communities.

In order to allow comparisons
with known results, we will: i) assume that the solution point
$ (\mathbf{x} , \theta) \in  Q $ is sufficiently removed from 
boundaries to allow neglect of boundary interaction effects,
and ii) focus on one-dimensional problems. The method to be 
described carries over without difficulty to multiple dimensions, and as noted,
can be  adapted to handle boundary conditions. 
In addition, for notational simplicity, we will
set $ \nu = 1 ; $ based on the discussion immediately above, this
corresponds to nondimensionalizing (\ref{general})
using $ l_b $ (or equivalently, $ l_D ) , $ $ b , $ and $ \Delta \theta $ for the length,
drift, and time scales, respectively, and then to setting $ Re_b = 1 . $

Thus, we express the time-differential form of (\ref{sol1}) in either of two equivalent forms:
\begin{equation}\label{difftl}
\eta (x,\theta_j) = \int_{-\infty}^{\infty} p(x, \theta_j | x',
\theta_{j+1}) ) \eta (x', \theta_{j+1}) dx' - E \int_{\theta_j}^{\theta_{j+1}} f (\chi(s'),s') ds'
\end{equation}
or
\begin{equation}\label{difftl2}
\eta (x,\theta_j) \approx  \sum_{i=1}^{N_x} \Big[ \frac{n(x_i^{'} , \theta_{j+1} )}{N} \Big]
\eta ( x_i^{'} , \theta_{j+1} ) 
- \frac{1}{N} \sum_{k=1}^N  \int_{\theta_j}^{\theta_{j+1}} f( \chi_k (s') , s') ds'
\end{equation}
where $ p $ is the transition density, $ n (x_i^{'} , \theta_{j+1} ) / N  $ is the 
fraction of $ N $ RW's launched from the solution point $ ( x , \theta_j ) $ 
reaching a line segment, $ \Delta x_i $ (on the x-axis), during the interval 
$ ( \theta_j , \theta_{j+1} ] , $ $ x_i^{'} \in \Delta x_i , $ and $ N_x $ is the number of 
increments used to discretize a length $ 2 L $ of the x-axis. 
Here, $ 2 L $ is, e.g., centered on $ x , $
and $ L $ is chosen
to be much larger than the maximum RW displacement produced by
combined advection and diffusion over $ \Delta \theta . $

With regard to the differential solutions in (\ref{difftl}) and (\ref{difftl2}),
we note that (\ref{difftl}) is useful when the transition density, $ p , $
can be determined analytically via the FP equation; 
this approach is illustrated in our derivation of the Cole-Hopf/KPZ (CH/KPZ) transformation
below. The second form, (\ref{difftl2}), which is useful, for example,
in Monte Carlo simulations where analytical $ p 's $ are not available,
is based on the following simple relationship between $ p $ and
the fraction $ n (x_i^{'} , \theta_{j+1}) / N : $
$ p (x , \theta_j | x_i^{'} , \theta_{j+1} ) = \lim_{\Delta x_i \rightarrow 0}  
n (x_i^{'} , \theta_{j+1}) / ( N \Delta x_i ) . $ 
As a practical aside, and considering (\ref{difftl2}),
the fractions $ n (x_i^{'} , \theta_{j+1}) / N $ are obtained by launching a swarm 
of $ N $ RW's from the solution point $ ( x, \theta_j ) , $ and then 
by counting the number 
of RW's
that subsequently reach each $ \Delta x_i $ over 
the increment $ ( \theta_j , \theta_{j+1}] . $

With regard to the path integral terms in (\ref{difftl}) and (\ref{difftl2}),
these can be expressed, depending on the application, as
$ f ( x , \theta_{j+1}) \Delta \theta + O ( \Delta \theta^2 ) , $ or
$ f ( x , \theta_{j}) \Delta \theta + O ( \Delta \theta^2 ) . $ 
This point is exploited in our derivation of
the Feynman-Kac formula below.

From a mathematical standpoint,
the power of both the incremental and
non-incremental Kolmogorov solution, (\ref{difftl}) (or (\ref{difftl2}))
and (\ref{sol1}), respectively,
derives from the fact that these
embody \textit{representation  theorems} [see, e.g., \cite{ma}]
by which, in the present context, \textit{any} entity,
$ \beta(\mathbf{p},\omega) , $
satisfying a generalized transport equation of the form 
$ \beta_{\omega} + \mathbf{v} \cdot \nabla_{\mathbf{p}}
\beta + \kappa \nabla_{\mathbf{p}}^2 \beta = F (\beta) ,  $
and having an associated stochastic differential
equation (SDE), $ d \boldsymbol{\chi} (\omega) = 
\mathbf{v} (\mathbf{p} , \omega) + \sqrt{2 \kappa} d \mathbf{w} (\omega) ,$
can be determined in time-incremental fashion.
Here, $ \mathbf{p} $ and $ \omega $ represent variables that are,
in the sense of this generalized transport equation, space- and time-like, respectively,
$ \mathbf{v} $ and $ \kappa $ are, respectively, drift field
and diffusion coefficient analogs,
$ F $ is a generalized source term, and $ \mathbf{w} $ is a Wiener process.

From a physical standpoint,
(\ref{difftl}) and (\ref{difftl2}) likewise embody powerful features.
In particular, since solutions of (\ref{general}) via (\ref{difftl}), or equivalently
(\ref{difftl2}),
require
determination of the transition density $ p $ (or equivalently, 
$ n (x_i^{'} , \theta_{j+1}) / ( N \Delta x_i ) ) , $ 
where $ p $ typically captures short time- and length-scale physics,
then in constructing solutions, we are immediately 
forced to confront physical features extant on these scales.
In many continuum-scale scalar transport problems, for example,
one can often reasonably assume that small scale
statistical features are adequately captured by the
Fokker-Planck, or somewhat more generally, the 
differential Chapman-Kolmogorov (CK) equation \cite{gardiner}.
In these cases, and again as illustrated in the first example below,
due to the local nature
of (\ref{difftl}) or (\ref{difftl2}), $ p $ typically
corresponds to a fundamental solution of the FP or CK equations.
Similarly, in quantum-scale problems, where (\ref{difftl}) (or (\ref{difftl2}))
corresponds to a time-incremental solution for a wave
function, $ \psi , $ via a propagator $ U $ (see, e.g., \cite{shankar})
(where $ \psi $ and $ U $ correspond respectively to
$ \eta $ and $ p ) , $
one must typically construct a Hamiltonian appropriate to the 
quantum system of interest.
%In situations where a key assumption underlying the FP or CK equation
%is violated, e.g., when stochastic features 
%are non-Markovian, then we must abandon
%the solution in (\ref{sol1}), or equivalently in (\ref{difftl}) and
%(\ref{difftl2}).

\vspace{.4cm}

\noindent \textbf{2. Example I: Cole-Hopf/KPZ transformation}  \\
In order to illustrate the utility of (\ref{difftl}) (or (\ref{difftl2}))
in solving nonlinear problems, we choose two relatively simple
embodiments of (\ref{general}), the noise-free Burger's and KPZ equations,
and show how the solution, (\ref{difftl}), encompasses, and can be used to derive,
the CH/KPZ
transformations. 
Importantly, this exercise exposes two significant features. First,
it illustrates a well known argument \cite{gardiner, schuss}
for obtaining a typical Gaussian transition density via solution of the FP equation;
again, determination of $ p , $ either analytically by solution of
the FP or CK equations, or by Monte Carlo simulation (leading to (\ref{difftl2}))
comprises the crucial step in solving (\ref{general}) via (\ref{difftl})
or (\ref{difftl2}). 

Second, and reflecting what appears to be a new result,
this exercise shows that in some cases, one can jump from
a time-incremental solution, as in (\ref{difftl}) or (\ref{difftl2}),
to a non-incremental solution, as in the CH/KPZ transform solutions.
The idea centers on circumventing a path integral,
$ E \int_{\theta_j}^{\theta_{j+1}} u (\chi(s'),s') ds' , $
over an \textit{a priori} unknown field, $ u , $
by first introducing a transformation $ \phi ( x', \theta ) = \exp ( f(u)) $
into the representation (\ref{difftl}), and then by insisting
that the stochastic construction of
$ \phi $ be purely diffusive, i.e., that the stochastic process, $ d \chi_{\phi} (s) , $
used to construct $ \phi $ be governed by $ d \chi_{\phi} (s) = \sqrt{2} d w(s) . $

Thus, considering the noise-free Burger's equation, 
$ u_{\theta} + u u_x + u_{xx} = 0, $
we set $ \eta = u $ 
in (\ref{difftl}),
where, for hydrodynamic problems, $ u $ is
the fluid velocity.
The solution in (\ref{difftl}) then takes the form
\begin{equation}\label{burgersdifftl}
u(x, \theta_j) = \int_{- \infty}^{\infty} u(x',\theta_{j+1}) 
p(x , \theta_j | x' , \theta_{j+1} ) dx' 
\end{equation}
This is a time-incremental, stochastic solution
to Burger's equation, and as far as we can determine,
has not been reported.

In order to determine $ p : $
i) we again assume that the solution point $ (\mathbf{x}, \theta ) $ is
far removed from both $ \delta Q $ and the final time slice $ D \times \{ \theta = T \} , $
ii) again choose the time increment $ \Delta \theta $
so that the local 
Reynolds number, or its equivalent, satisfies
$ Re_b = O(1) , $ and iii) set the source term $ f $ in (\ref{general}) to $ 0. $ 
Under these conditions,
the transition density, 
$ p(x, \theta_j | x' , \theta_{j+1}) , $ 
can be determined explicitly using
the backward Fokker-Planck equation,
$ p_{\theta} + u (x' , \theta) p_{x'} + p_{x'x'} = 0 , $
subject to the condition 
$ p(x, \theta_j | x' , \theta_j) = \delta ( x - x' ) , $  with the 
result \cite{gardiner, schuss}: 
$ p(x, \theta_j | x' , \theta_{j+1}) = [\sqrt{4 \pi \Delta \theta}]^{-1} 
\exp -[(x'- \bar{x})^2/(4 \Delta \theta)] , $
where again, $ \bar{x} = x + E \int_{\theta_j}^{\theta_{j+1}} b(\chi(s') , s') d s' $
is the average position of the RW swarm at $ \theta_{j+1} . $ 
Again, since 
$ E \int_{\theta_j}^{\theta_{j+1}} u (\chi(s') , s' ) ds' = u (x, \theta_j ) \Delta \theta 
+ O ( \Delta \theta^2 ), $ 
the presence of the 
path integral 
does not prevent time-incremental solutions;
various strategies, involving, e.g.,
Taylor expansions relating the unknown $ u (x, \theta_j ) $ to known 
$ u ( x, \theta_{j+1} ) , $ can be easily 
envisioned. 

Next, arriving at the crucial step in the
derivation of the Cole-Hopf solution, we take advantage
of the general nature of (\ref{difftl}), and as mentioned above,
introduce a transformation  
\begin{equation}\label{transform}
\phi (u) = \exp ( f(u) ) 
\end{equation}
into (\ref{difftl}). Again, in order to circumvent
the path integral over unknown $ u , $
we force the stochastic construction of $ \phi $ to be
purely
diffusive, $ \phi_{\theta} + \phi_{xx} = 0 . $
Importantly, defining a purely diffusive $ \phi $ of the form given in (\ref{transform})
allows us, on one hand,
to stretch the solution interval to any arbitrary length (by eliminating the path integral),
while on the other, providing us with 
sufficient freedom to identify transformations, $ f(u) , $
between the simple linear heat equation governing $ \phi , $ and more complicated, nonlinear
equations governing $ u . $
It appears likely that
a similar strategy can be used to find long-time (non-incremental)
transformations and solutions for other nonlinear equations.

Thus, recognizing that $ p $ is now governed by
$ p_{\theta} + p_{xx} = 0 , $
and inserting $ \phi $ in (\ref{difftl}), we obtain
\begin{equation}\label{phieqn}
\phi(x, \theta_j) = \int_{-\infty}^{\infty} \phi(x', \theta_{j+1} ) 
\Big[ \sqrt{4 \pi (\theta_{j+1} - \theta_j)} \Big]^{-1} 
\exp -[(x'- x)^2/(4 (\theta_{j+1}-\theta_j) )] dx'
\end{equation}
where now $ \bar{x} = x . $

At this stage, several equivalent paths can be taken to
complete the derivation.
The most transparent, which also provides the KPZ transform,
first determines $ f (u) $ by finding an equivalent transformation,
$ \phi = \phi (\tilde{f}(v)) , $ that solves the related \textit{forward-time}, 
noise-free KPZ equation,
$ v_t + v_x^2 /2 - v_{xx} =0 . $ This is 
followed by use of the well-known transformation \cite{fogedby},
$ u = v_x , $ between the KPZ and Burger's equations, after which
$ \phi $ is introduced into (\ref{difftl}), $ \theta_{j+1} $ stretched to
the final backward time $ T , $ and the final result obtained.

Thus, the forward-time form of the governing equation for $ \phi , $ $ \phi_t - \phi_{xx} = 0 , $
yields $ v_t - [ \tilde{f}_v + \tilde{f}_{vv}/\tilde{f}_v] v_x^2 - v_{xx} = 0 , $
which, in turn, transforms to the 
KPZ equation by 
choosing $ [\tilde{f}_v + \tilde{f}_{vv}/\tilde{f}_v] = - 1/2 $ (where we again set the
coefficient multiplying $ v_{xx} $ to 1). 
Note, we change to a forward time solution for $ \phi $ in order to obtain
the correct signs on $ v . $ This is acceptable (and an advantage of this approach) 
since the purely diffusive stochastic
process, $ \chi_{\phi} , $ used to construct $ \phi $
can be run in either the forward or backward time direction.
In order to
obtain a typical form of 
the KPZ transformation and solution \cite{fogedby}, we choose $ \tilde{f}_v  = - 1/2 , $
which then yields $ \phi = \exp (-v/2) , $ or $ v(x,t) = -2 \ln { \phi } , $
identical to the form given in \cite{fogedby} (with $ \nu = 1 ) . $

Next,  
restate $ \phi $ 
in terms of $ u , $ $ \phi = \exp { \big[ - 1/2 \int^x u dx \big] } , $
and insert this into (\ref{difftl}). 
Then isolate 
$ u(x, \theta_j) $ using $ u (x , \theta_j ) = -2 \phi_x / \phi |_{x,\theta_j} , $ 
set $ \theta_j = t , $ $ \theta_{j+1} = T , $ and
finally obtain, after some manipulation,
the classic Cole-Hopf solution of Burger's equation (see, e.g., \cite{whitham} ):
\begin{equation}\label{burgers}
u(x , t) = \frac{ \int_{-\infty}^{\infty} \frac{(x-x')}{t} \exp (-G/2) dx'}
{ \int_{-\infty}^{\infty} \exp (-G/2) dx'}
\end{equation}
where $ G = G (x';,x, t) = \int_0^{x'} u(y,t=0) dy + \frac{(x-x')^2}{2t} . $

\vspace{.4cm}

\noindent \textbf{3. Example II: Feynman-Kac formula} \\ 
Turning next to solution of nonhomogeneous versions of (\ref{general})
via (\ref{difftl}),
we illustrate by using (\ref{difftl}) to derive
the Feynman-Kac formula.
In particular, consider a one-dimensional Schr\"odinger-like transport equation of the form
$ \eta_{\theta} + b \eta_x + \nu \eta_{xx} = V(x,\theta) \eta , $
where $ b $ and $ \nu $ either correspond to, or are 
local analogs to drift and diffusivity, and where $ V $ is a deterministic function.
The Kolmogorov solution (\ref{difftl}) over $ \Delta \theta $ is thus stated as
\begin{equation}\label{kac}
\eta (x,\theta_j) = \int_{-\infty}^{\infty} p(x, \theta_j | x',
\theta_{j+1}) ) \eta (x', \theta_{j+1}) dx' 
- E \int_{\theta_j}^{\theta_{j+1}} V (\chi(s'),s') \eta ( \chi(s') , s') ds' . 
\end{equation}

For present purposes, there is no need to specify $ p . $
Thus, we first note that 
$ E \int_{\theta_j}^{\theta_{j+1}} V (\chi(s'),s') \eta ( \chi(s'), s') ds' = 
V(x, \theta_j) \eta( x, \theta_j ) \Delta \theta
+ O (\Delta \theta^ 2) . $  Hence, to $ O(\Delta \theta^2 ) , $
(\ref{kac}) can be expressed as \\
$ \eta(x ,\theta_j) = (1 + V \Delta \theta)^{-1}
\int_{-\infty}^{\infty} p(x, \theta_j | x',
\theta_{j+1}) ) \eta (x', \theta_{j+1}) dx' , $ or, using $ ( 1 + V \Delta \theta )^{-1} =
(1 - V \Delta \theta ) + O(\Delta \theta^2 ) = 
\exp[ - V( x , \theta_j ) \Delta \theta ] + O (\Delta \theta^2 ) , $
as $ \eta(x ,\theta_j) = 
\exp[ - V( x , \theta_j ) \Delta \theta ] 
\int_{-\infty}^{\infty} p(x, \theta_j | x',
\theta_{j+1}) ) \eta (x', \theta_{j+1}) dx' 
+ O (\Delta \theta^2 )  . $

Next, taking the expectation and suppressing
the error estimate, we rewrite the preceding expression
for $ \eta(x , \theta_j ) $ as 
$ \eta(x ,\theta_j) = 
\exp[ - V( x , \theta_j ) \Delta \theta ] \eta(\bar{x}_{j+1} , \theta_{j+1}) , $ 
where the average position
of the RW swarm at $ \theta_{j+1} $ is expressed as $ \bar{x}_{j+1} . $ 
With this last expression,
we can now connect $ \eta (x , \theta_j ) $
to $ \eta (\bar{x}_N , \theta_N ) , $ the value of $ \eta $
in the final time-slice $ D \times \{\theta=T \} , $
evaluated at the swarm's final mean position, $ \bar{x}_N  ; $
thus, $ \eta (x , \theta_j ) = \exp \Big[ - \Delta \theta 
\big[ V (x , \theta_{j} ) + V (\bar{x}_{j+1} , \theta_{j+1} )
+ V ( \bar{x}_{j+2} , \theta_{j+2} ) + \ldots + V (\bar{x}_N , \theta_N ) \big] \Big] 
\eta (\bar{x}_N , \theta_N ) , $
or finally, 
$ \eta (x , \theta_j ) = E \Big[ \exp \big[ - \int_{\theta_j}^{\theta_{N}} 
V (\chi(s') , s') ds' \big] \Big]
\eta(\bar{x}_N, \theta_N ) . $ 

A closing note regarding nonhomogeneous problems: as before, in cases where
an analytical solution for the transition density, $ p , $ cannot be obtained,
then an incremental stochastic construction can be carried out using an appropriate SDE
combined with the approximate representation in (\ref{difftl2}).
In the present illustration, for example, the stochastic processes
used to construct $ \eta $ are governed by $ d \chi (\theta) =
b( \chi (\theta) , \theta ) d \theta + \sqrt{2 \nu} d w(\theta) . $

\end{document}